\documentclass[journal]{IEEEtran}
\usepackage{cite}
\usepackage{amsmath,amssymb,amsfonts}
\usepackage{graphicx}
\usepackage{subfig}
\usepackage[algo2e]{algorithm2e}
\usepackage[ruled]{algorithm}
\usepackage{tikz}
\usetikzlibrary{calc}
\usepackage{colortbl}
\usepackage{float}
\usepackage{threeparttable}
\usepackage{tabularx}
\usepackage{multirow} 
\usepackage{booktabs}
\usepackage{url}
\usepackage{hyperref}
\usepackage{enumitem}
\setlist[itemize]{left=0pt} 

\def\BibTeX{{\rm B\kern-.05em{\sc i\kern-.025em b}\kern-.08em
T\kern-.1667em\lower.7ex\hbox{E}\kern-.125emX}}

\begin{document}
\title{Semi-Supervised Learning for Dose Prediction in Targeted Radionuclide: A Synthetic Data Study}



\author{Jing Zhang, Alexandre Bousse, Chi-Hieu Pham, Kuangyu Shi, and Julien Bert

\thanks{This work did not involve human subjects or animals in its research.}

\thanks{This work was partially funded by the European Union through the SECURE project under the grant HORIZON-EURATOM-2021 N$^o$ 101061230 and by the French National Research Agency with the project MoCaMed under reference ANR-20-CE45-0025.}
\thanks{The authors are with LaTIM, INSERM-UMR1101, University of Brest, 29200 Brest, France (e-mail: jing.zhang@univ-brest.fr). K. Shi are with University of Bern, Switzerland.}
}
\maketitle
\begin{abstract}
\bfseries 
Accurate and personalized radiation dose estimation is crucial for effective Targeted Radionuclide Therapy (TRT). Deep learning (DL) holds promise for this purpose. However, current DL-based dosimetry methods require large-scale supervised data, which is scarce in clinical practice. To address this challenge, we propose exploring semi-supervised learning (SSL) framework that leverages readily available pre-therapy PET data—where only a small subset requires dose labels—to predict radiation doses, thereby reducing the dependency on extensive labeled datasets.
In this study, traditional classification-based SSL approaches were adapted and extended in regression task specifically designed for dose prediction. To facilitate comprehensive testing and validation, we developed a synthetic dataset that simulates PET images and dose calculation using Monte Carlo simulations.
In the experiment, several regression-adapted SSL methods were compared and evaluated under varying proportions of labeled data in the training set. The overall mean absolute percentage error of dose prediction remained between 9\% and 11\% across different organs, which achieved comparable performance than fully supervised ones. 
The preliminary experimental results demonstrated that the proposed SSL methods yield promising outcomes for organ-level dose prediction, particularly in scenarios where clinical data are not available in sufficient quantities.
\end{abstract}


\section{Introduction}\label{sec:introduction}

\IEEEPARstart{T}{argeted} Radionuclide Therapy (TRT) is a contemporary approach in radiation oncology, aiming to deliver therapeutic radiation doses using cancer-targeting radiopharmaceuticals. Despite the early success of TRT, concerns have been raised about the risks of an inadequate trade-off between therapeutic dose and side effects \cite{sgouros2020radiopharmaceutical}. An essential requirement of TRT is optimizing the radiation dose adapted for individual patients. One idea is to predict the absorbed dose in advance of therapy \cite{sgouros_icru_2021}. For this, one possibility is the use of Monte Carlo (MC) simulation which is considered as the gold standard for dose calculation in medical physics due to its ability to model complex interactions of radiation with matter. However, its high computational cost makes it difficult to use in a clinical context. As a result, researchers are exploring alternative methods, such as deterministic algorithms including the dose point kernel \cite{giap1995dpk} and the voxel S-value \cite{lee2018vsv}, to achieve a balance between accuracy and efficiency.

Deep Learning (DL) approaches may play a key role in personalizing the dose in TRT using Positron Emission Tomography (PET) pre-therapy imaging  \cite{xue2024prepet,xue2023dl}. However, DL requires access to a large amount of training data, which is not easy to obtain. In TRT, pre-therapy PET images predominantly use the same radioisotope, which facilitates the accumulation of training data in quantities suitable for deep learning (DL) approaches. However, this is not applicable to dosimetry data derived from SPECT images captured at various points during the treatment cycle. These images are highly dependent on the specific treatment radionuclide used and are available in smaller quantities. For instance, for the prostate-specific membrane antigen (PSMA) in prostate cancer, both treatments with $^{177}$Lu-PSMA (beta) and with $^{225}$Ac-PSMA (alpha) \cite{parker2018alpha} share the same pre-therapy imaging protocol with $^{68}$Ga-PSMA, but the SPECT images and dosimetry data will differ. This highlights that pre-therapy PET data is more readily available in larger quantities, especially for already established TRT, compared to dosimetry data, which is available in smaller quantities, particularly when implementing innovative radioisotopes. Therefore, training DL models may be challenging \cite{strigari2025computational} and rises significant obstacles to developing and validating robust DL solutions for treatment planning and dosimetry.

To address these challenges, this work explores semi-supervised learning (SSL) approaches for the first time to enable personalized patient dosimetry prediction from pre-therapy images. The aim of SSL is to leverage large amounts of unlabeled data (PET pre-therapy without post-therapy dosimetry) to enhance dosimetry of TRT involving radioisotopes under investigation and deployment where only small sets of labeled data (post-therapy dosimetry) is available. One limitation, of course, is that the targeted molecule in the unlabeled data (e.g., PSMA) must be the same as that in the labeled data. Otherwise, the biodistribution will differ. For instance, large amounts of PET data from common protocols, such as $^{68}$Ga-PSMA can be used to improve dose prediction for $^{225}$Ac-PSMA, where clinical data is limited. To facilitate comprehensive testing and validation, we developed a synthetic dataset that simulates PET images and dose calculation using MC, enabling extensive experimentation across various clinical scenarios and SSL models. The proposed work and study on synthetic data focus solely on organs at risk, excluding the dose to metastatic tumors. Furthermore, traditional classification-based SSL approaches was extended in regression task using a new Pseudo-Label loss specifically designed for TRT dose prediction.

\section{Related works}

\subsection{Dose prediction with fully supervised learning}

In the context of dosimetry, the labeled data refers to the attenuation map as well as the corresponding dose value computed by MC simulation. In DL domain, dose prediction corresponds to a regression task, various techniques for dose prediction based on DL began to emerge increasingly \cite{visvikis2022application}. In internal dosimetry, Lee et al. \cite{lee2019dosedl} used a U-Net \cite{ronneberger2015unet} to predict the absorbed dose from PET/CT images. Although pre-therapy PET captures only a single time point and differs from therapy such as $^{177}$Lu in half-life and pharmacokinetics, introducing inherent uncertainty, pre-therapy PET (e.g., $^{68}$Ga-PSMA) has been explored as a surrogate for $^{177}$Lu-PSMA dosimetry \cite{peters202268ga}. Similarly, Xue et al. \cite{xue2023dl} used generative adversarial network (GAN) \cite{goodfellow2020generative} with a U-Net generator and a convolutional neural network (CNN) based discriminator on dose estimation. In TRT, Mansouri et al. \cite{mansouri2024vitdose} used a vision transformer model \cite{dosovitskiy2020image} with a Multi-Head Attention mechanism \cite{vaswani2017attention} to predict the post-therapy dose of a $^{177}\text{Lu}$-PSMA-I\&T radiopharmaceutical therapy using multiple time points of SPECT images. In addition, single-time-point dosimetry offers a promising approach to further simplify personalized dosimetry while maintaining accuracy \cite{sadeghi2025current,wang2023single}. DL approaches were also used in standard radiotherapy treatment, such as in intensity-modulated radiation therapy \cite{nguyen2019dosedl} or in prostate brachytherapy \cite{villa2022}. Recent architectures have also been explored, such as Diffusion models \cite{ho2020ddpm}. For example, in the study by \cite{zhang2024dosediff} on breast cancer and nasopharyngeal cancer, and the study by \cite{fu2024mddose} on thoracic tumor patients, dose prediction was defined as a sequence of denoising steps. Another study by \cite{maniscalco2024multimodaldose} demonstrated the advantages of a multi-task DL framework for predicting dose distributions across different image modalities compared to single-task models. 

Previous studies have primarily relied on fully supervised DL methods, despite the limited availability of clinical data. While these approaches have achieved promising accuracy and efficiency in dose estimation tasks, they are also prone to overfitting when the amount of training data is insufficient relative to the model’s complexity. To mitigate these limitations, semi-supervised learning has emerged as a promising alternative \cite{yang2022sslsurvey}, as it can effectively leverage both labeled and unlabeled data to improve generalization while reducing reliance on large-scale annotated datasets.

\subsection{Semi-supervised learning}
\label{sec:sotassl}
Among the various training strategies in SSL, three major paradigms have received significant attention: consistency regularization, pseudo-labeling, and hybrid models. Consistency regularization is a widely used strategy in SSL, where the model is encouraged to produce consistent predictions under input perturbations. It is commonly implemented in a Teacher-Student framework \cite{tarvainen2017meanteacher-ssl, laine2016temporal-ssl, xie2020uda-ssl}, where the Teacher model provides stable targets to guide the learning of the Student model. Unlike the Student model, the parameters of the Teacher model are not updated using gradient-based optimization but are instead adjusted through exponential moving average (EMA) to ensure stability. This approach enhances generalization by leveraging unlabeled data, but may suffer from hyperparameter sensitivity and confirmation bias \cite{yang2022sslsurvey}. 

Pseudo-labeling is another SSL approach that assigns labels to unlabeled data based on the model’s high-confidence predictions \cite{lee2013pseudo}. By incorporating pseudo-labeled samples into the training set, the model can benefit from additional supervision and achieve better generalization \cite{mcclosky2006selftraining-ssl, blum1998cotraining-ssl}. However, generated pseudo labels may be noisy, particularly during the early training stages, leading to confirmation bias and error reinforcement. Moreover, its performance is often sensitive to the distribution of labeled data. Thus, careful selection of pseudo-labeled samples is crucial, for example, by adopting curriculum strategies \cite{cascante2021curriculumlabel-ssl}. Lastly, hybrid models \cite{berthelot2019mixmatch} effectively addressed the limitations of both consistency regularization and pseudo-labeling by capitalizing on their respective strengths. Consistency regularization promoted the acquisition of generalizable features, while pseudo-labeling method enabled the utilization of unlabeled data. This synergistic combination empowered hybrid models to achieve superior performance across a broader spectrum of tasks, for instance with the convolutional networks based FixMatch \cite{sohn2020fixmatch-ssl} and attention-based hybrid model \cite{cai2022semivit-ssl}. 

While the aforementioned SSL models have demonstrated success in classification tasks, they exhibit limitations when applied to regression tasks. In particular, confidence-based pseudo-labeling methods, which rely on high-confidence predictions for generating pseudo labels, are not directly applicable to regression problems due to the continuous nature of the output. There are currently no studies that involve the use of SSL in dose calculation, particularly none dedicated to TRT. Therefore, this work explores, for the first time, how these models can be adapted and extended for regression tasks, specifically for TRT dose calculation.

\begin{figure}[t]
    \centering
    \includegraphics[width=0.5\textwidth]{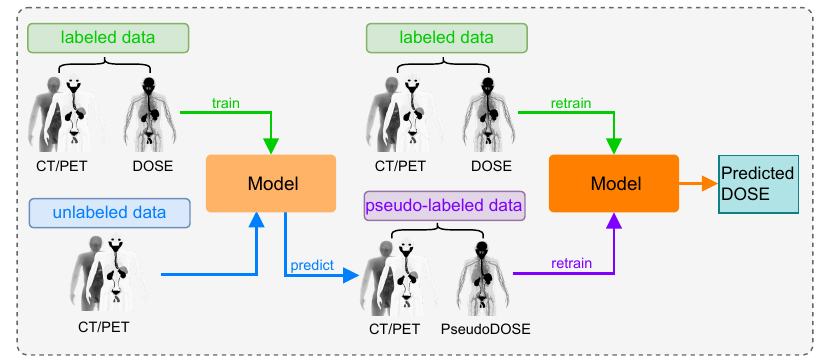}
    \caption{Overview of the proposed semi-supervised learning framework for dose prediction on a synthetic dataset. Limited labeled data (PET with dose) were used to initialize the model, while unlabeled data (only PET) were iteratively assigned pseudo-labels and incorporated into training.}

    \label{fig:flowchart}
\end{figure}

\section{Materials and Methods}

The complete workflow of our SSL framework is illustrated in Fig.~\ref{fig:flowchart}. Since the paired PET and dose data are scarce in real clinical practice, we employed a synthetic dataset to both ensure reproducibility and simulate this realistic limitation. In our semi-supervised setting, a small subset of samples with ground-truth dose distributions was treated as labeled data, while the majority was considered unlabeled. The model was initially trained with the labeled subset and then applied to the unlabeled data to generate pseudo-labels. These pseudo-labeled samples were incorporated into the training set, so that the model is retrained on both labeled and pseudo-labeled data in an iterative manner. This procedure reflects the central idea of semi-supervised learning: leveraging abundant unlabeled imaging data to progressively expand the effective training set and improve dose prediction accuracy, even when only limited labeled data are available.

\subsection{Dose prediction with semi-supervised learning}
\label{sec:sslm}
When labeled data is scarce, SSL approaches extends the conventional supervised learning paradigm by incorporating both supervised and unsupervised losses into the optimization process of learnable parameters ($\theta$) in SSL models. The overall objective can be formulated as follows:

\begin{equation}
\min_{\theta} \underbrace{\mathcal{L}_s(\lambda_{lb}, \theta)}_{\text{supervised loss}} + \alpha \underbrace{\mathcal{L}_u(\lambda_{ulb}, \theta)}_{\text{unsupervised loss}}
\label{eq:ssl}
\end{equation}

\noindent where $\lambda_{lb}$ and $\lambda_{ulb}$ denote labeled and unlabeled images, respectively. The supervised loss $\mathcal{L}_s$ ensures proper guidance from labeled data, while the unsupervised loss $\mathcal{L}_u$ can take various forms, depending on the specific SSL strategy.

\begin{figure}[t]
\centering
    \subfloat[MT]{\includegraphics[width=4.5cm]{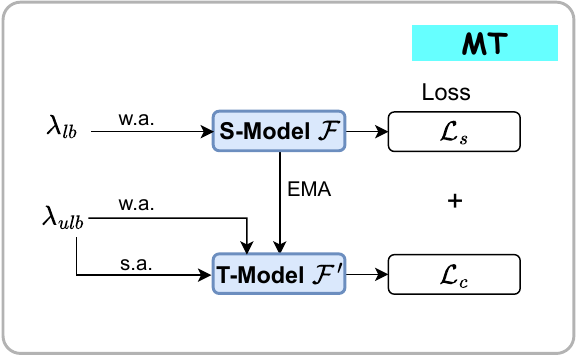}}~
    \subfloat[ICT]{\includegraphics[width=4.5cm]{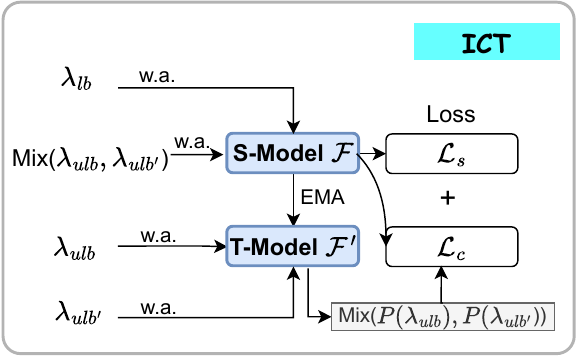}}
    
    \subfloat[FixMatch]{\includegraphics[width=4.5cm]{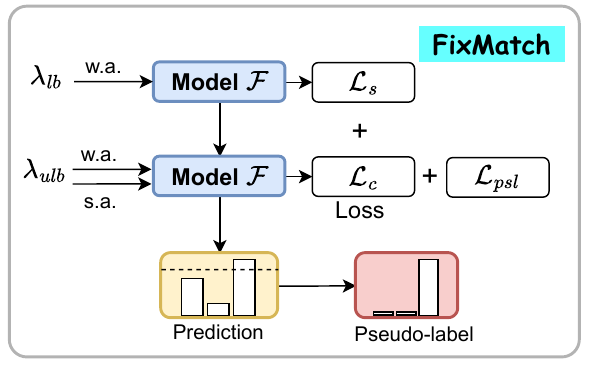}}~
    \subfloat[SGAN]{\includegraphics[width=4.5cm]{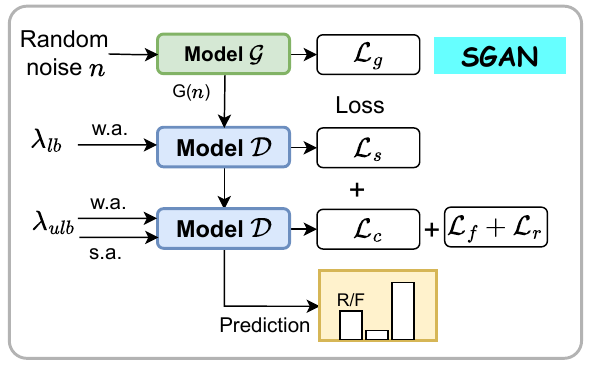}}
    \caption{SSL architectures for classification. (a) and (b) consistency regularization ($\mathcal{L}_c$); (c) pseudo-labeling method; (d) generative models. Weak and strong augmentation (w.a. and s.a.) are performed on the input labeled ($\lambda_{lb}$) and unlabeled ($\lambda_{ulb}$) images.}
    \label{fig:ssl-models} 
\end{figure}

In this work, we implemented several representative SSL frameworks, the training processes of these SSL models are illustrated in Fig. \ref{fig:ssl-models}. Mean Teacher (MT) \cite{tarvainen2017meanteacher-ssl} and Interpolation Consistency Training (ICT) \cite{verma2022ict} shown in Fig. \ref{fig:ssl-models}(a-b) rely solely on consistency regularization, which enforces the model to produce similar predictions for perturbed versions of the same input, i.e., $\mathcal{L}_u(\lambda_{ulb}, \theta) = \mathcal{L}_c$. Student model was trained with labeled data ($\lambda_{lb}$); unlabeled data ($\lambda_{ulb}$) was trained on Teacher model;  FixMatch \cite{sohn2020fixmatch-ssl} (see Fig. \ref{fig:ssl-models}(c)) further combines consistency loss with pseudo-labeling method, denoted as $\mathcal{L}_u(\lambda_{ulb}, \theta) = \mathcal{L}_c + \mathcal{L}_{psl}$, where high-confidence predictions on weakly augmented unlabeled data are treated as labels for strongly augmented versions of the same data. Generative SSL models such as SGAN \cite{toutouh2023semigan} in Fig. \ref{fig:ssl-models}(d) incorporate both consistency loss and adversarial loss to distinguish real from fake images. 

\begin{figure}[t]
    \centering
    \includegraphics[width=8.2cm]{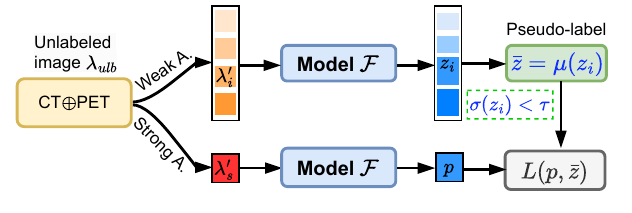}
    \caption{\small{Diagram for the proposed Pseudo-Label generation in SSL for regression. The model is trained on a set of weakly augmented data ($\lambda'_{i}$) from unlabeled data ($\lambda_{ulb}$). $\bar{z}$ is pseudo-label. A predicted dose value ($p$) is then inferred from strongly augmented data ($\lambda'_{s}$). Finally, the model is updated using the regression loss $L(p,\bar{z})$.}}
    \label{fig:ssl}
\end{figure}

\subsection{Pseudo-Label loss (PSL) for regression}
While pseudo-labeling method is effective for classification tasks, this approach is inherently unsuitable for regression problems, as it relies on discrete label assignments rather than continuous value estimation.  To address this, we proposed a Pseudo-Label loss tailored for regression tasks. As Fig. \ref{fig:ssl} illustrated, given a model $\mathcal{F}$ with learnable parameters $\theta$, we first obtain multiple predictions $z_i$ from weakly augmented input $\lambda'_i$ (e.g., small rotations, Gaussian noise), as follows: 

\begin{equation}
    z_i = \mathcal{F}_{\theta}(\lambda'_i)
    \label{eq:ps}
\end{equation}

If the predictions $z_i$ show low variance (i.e., $\sigma(z_i) < \tau$), it suggests that the model is confident about its prediction for this input. In this case, we take the average $\bar{z}$ as a pseudo-label. The model is then optimized by minimizing the discrepancy between $\bar{z}$ and the prediction obtained from a strongly augmented version (e.g., rotation, Gaussian noise, flipping, contrast adjustment, etc.) of the input $\lambda'_s$, using a regression loss function, shown as below:

\begin{equation}
    \min_\theta[L(\mathcal{F}_{\theta}(\lambda'_s), \Bar{z}(\sigma(z_i)<\tau))]
    \label{eq:loss}
\end{equation}

\begin{figure*}[t]
    \centering
    \includegraphics[width=15cm]{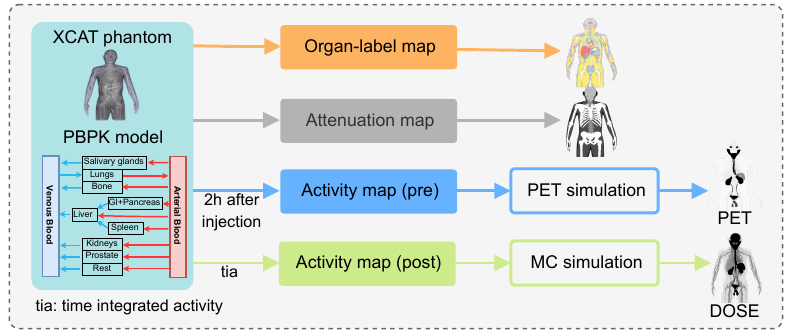}
    \caption{Schematic workflow of data generation process. A customized phantom is generated using XCAT to create an attenuation map and an organ-label map. Patient-specific pre- and post-therapy radiopharmaceutical activity maps are generated using the PBPK model. PET and MC simulations are then performed to obtain PET and dose maps.}
    \label{fig:flowchart_phantom}
\end{figure*}

Therefore, for dose prediction in regression SSL, the overall training consists of two parts: (1) an initial supervised training phase using labeled data, ensuring that the model learns a reasonable mapping from input to output, and (2) a refinement phase incorporating unsupervised loss such as consistency regularization and the proposed Pseudo-Label loss to further improve prediction accuracy. Notably, the pseudo-label generation module is designed to be model-agnostic, i.e., it can be seamlessly integrated into various SSL frameworks for regression tasks. In this work, we added the Pseudo-Label loss in the model: MT, ICT and SGAN, these SSL models are consistency learning based. For FixMatch, we replaced the original pseudo-labeling method only used for classification with our regression-based method, we call this model RegFixMatch.

\subsection{SynDoseTRT dataset}
The evaluation of SSL approaches requires comparing results with different amounts of data (various portions of the data). For this purpose, a large amount of data is required. Therefore, \textit{SynDoseTRT}, a synthetic dataset specifically designed for PET imaging and dose calculation in targeted radionuclide therapy was proposed. In this work, the dataset focused on prostate cancer ($^{177}\text{Lu}$-PSMA-I\&T), which is a common target for TRT and for which we have clinical data allowing us to design our synthetic data. The workflow of data generation (see Fig. \ref{fig:flowchart_phantom}) integrates anatomical phantoms, physiologically based pharmacokinetic (PBPK) modeling, PET simulation, and MC simulation to generate realistic imaging and dosimetry data. The detailed information of the final dataset is presented in Table \ref{tab:phantoms}, and the method by which it was obtained will be described in the subsequent sections.

\begin{table}[htp]
\centering
\caption{SynDoseTRT dataset information}
\label{tab:phantoms}
\begin{tabularx}{0.45\textwidth}{p{0.25\textwidth}p{0.2\textwidth}} 
\hline
Item & Value \\
\hline
Number of phantoms   & 1000 \\
Height & $ 173 \pm 4.4 $ cm \\
Weight & $ 74 \pm 2.8 $ Kg \\
Input PET radionuclide & $^{68}\text{Ga}$-PSMA-11 \\
Input PET dose & $ 115 \pm 10$ MBq \\
Input therapy radionuclide & $^{177}\text{Lu}$-PSMA-I\&T \\
Input therapy dose & $ 7400 \pm 150$ MBq \\
Pharmacokinetics model & PBPK \\
Phantom size & 256$\times$256$\times$256 \\
Voxel size & 3$\times$3$\times$4.5 mm$^3$ \\
\hline
\end{tabularx}
\end{table}

\subsubsection{Phantoms generation}

\begin{figure}[t]
\centering

\subfloat[Organ-label map]{\includegraphics[width=3.30cm]{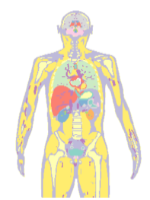}}~~
\subfloat[Attenuation map (MIP)]{\includegraphics[width=3.80cm]{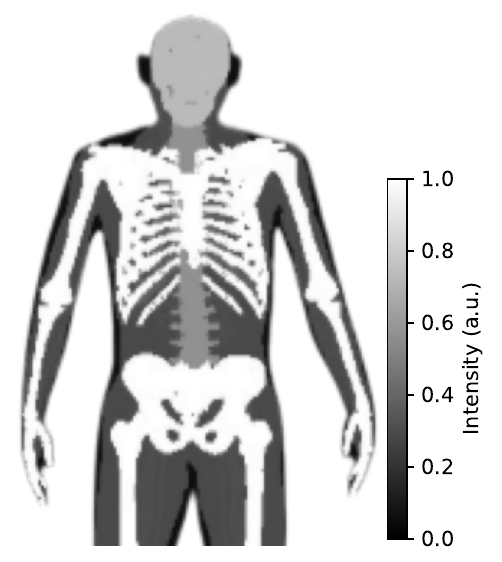}}

\subfloat[PET image at 2h time point (AIP)]{\includegraphics[width=3.80cm]{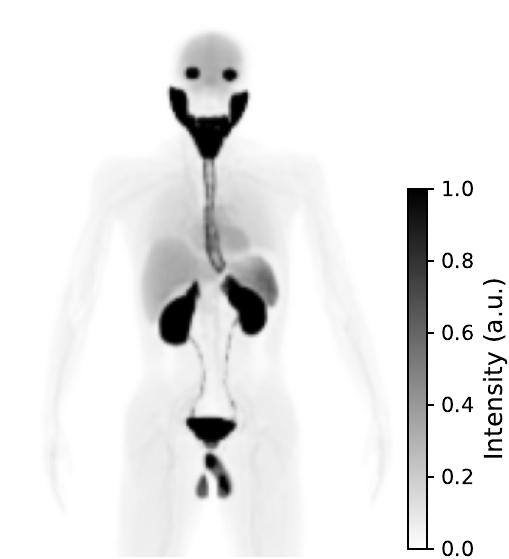}}~~ 
\subfloat[Dose image of TIA over 20 days (AIP)]{\includegraphics[width=3.8cm]{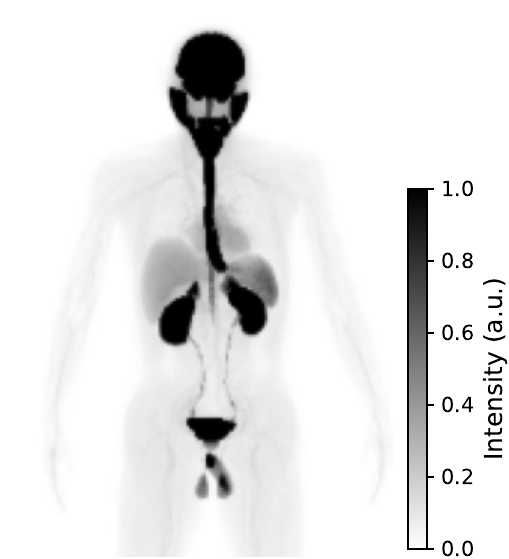}}  
\caption{One sample of synthetic phantom in SynDoseTRT dataset. (a) coronal slice organ-label map; (b) attenuation map; (c) coronal average projection PET; (d) coronal average projection dose map.}
\label{fig:phantom} 
\end{figure}

The phantom was generated using the XCAT \cite{segars2010xcat}, which is an anthropomorphic phantom model derived from real human anatomical data. XCAT phantom provides clear organ boundaries and detailed representations of human anatomy. In the SynDoseTRT dataset, the morphology of the phantom was randomly sampled to encompass various patient sizes and weights. Each XCAT phantom was voxelized, with each voxel assigned to a specific number representing organs and tissues, as well as corresponding attenuation coefficient values, an example is shown in Fig. \ref{fig:phantom}(a-b). The 3D attenuation and organ-label maps generated from XCAT will be used for both imaging and dosimetry simulations.

\subsubsection{PBPK model}

The PBPK model simulates the uptake, distribution, metabolism, and excretion of radiopharmaceuticals across different organs, providing time-activity curves (TACs) for each organ \cite{begum2019PBPK}. In this model, the human body is represented as multiple compartments, where each compartment corresponds to an organ or tissue, such as blood, liver, or kidneys, etc. The bloodstream transports the drug between different organs, while organs metabolize or eliminate the drug. Each of these processes evolves dynamically over time. These phenomena follow a system of differential equations, expressed as:

\begin{equation}
\frac{dC}{dt} = f(C, t, \Phi)
\label{eq:pbpk}
\end{equation}

\noindent where $C$ represents the drug concentration, $t$ denotes time, and $\Phi$ are the physiological parameters of the PBPK model. Ordinary differential equations ($f$) are used to describe the time-dependent variation of drug concentration. The parameters in PBPK model were initialized with organ-specific pharmacokinetic parameters derived from literature \cite{kassar2024pbpk} and clinical data.
 
In this work $^{68}$Ga-PSMA-11 pre-therapy PET imaging was considered. Then, by following the clinical protocol, the PBPK model was used to calculate the TAC of organs for 2 hours after injection. The total injected activity for a phantom was set to 115 ± 10 MBq, determined using weight-based dose protocol. The resulting organ-specific activity concentrations were mapped into a voxelized phantom for PET simulation. For post-therapy dosimetry estimation, we used a common PBPK structural model for both $^{68}\text{Ga}$-PSMA-11 and $^{177}\text{Lu}$-PSMA-I\&T for computational simplicity. We note that their in‑vivo biodistributions are not identical, and our use of a shared structure does not imply parameter equivalence. The PBPK model calculated time-integrated activity (TIA) by integrating the TACs over the full treatment period. The injected activity was set to 7400 ± 150 MBq, where the variation is based on the weights of phantoms. The calculated organ-specific TIA values were mapped onto the voxelized phantom, which was later used for dose calculation.

\subsubsection{PET simulation}

The pre-therapy PET simulation framework was implemented through physics-aware modeling as well as iterative reconstruction. Since Monte Carlo (MC) simulations are computationally demanding for generating large amounts of data, we instead used a fast PET simulation with a deterministic model. First, raw data projections were built using the activity and attenuation maps of the phantom, and then we applied iterative reconstruction method to obtain the final 3D PET image, see an example in Fig. \ref{fig:phantom}(c). This enables the simulation of PET images with realistic noise and resolution. In detail, raw data projections were built using parallel forward projections over $[0, 2\pi]$ with angular increments. Each projection integrates the spatial resolution of the detector with a Gaussian filter, phantom attenuation, and projection count with temporal scaling to simulate acquisition time. Finally, Poisson noise is added to the final projection to replicate photon counting statistics. Additionally, a constant background term accounting for scatter and random coincidences is added. The reconstruction process used a 100-iteration Maximum Likelihood Expectation Maximization (MLEM) algorithm \cite{shepp1982maximum} initialized with a uniform activity estimate. In terms of implementation, since our framework was developed in Python, the computations were accelerated using GPU through the CuPy library \cite{nishino2017cupy}.

\subsubsection{Absorbed dose calculation}

To determine the total therapeutic dose, MC simulation was performed using GATE v10~\cite{sarrut2021mc}. The simulation involves using the organ-label phantom and its corresponding material values for different tissues, along with the TIA map derived from the PBPK model, as well as the $^{177}$Lu radioisotope source. The 3D dose map was estimated using a sufficient number of particles (an not the total activity) to achieve an average dose uncertainty of less than 5\%. Afterwards, organ-specific absorbed doses (prostate, salivary glands, liver, spleen, pancreas, kidneys, bladder, and rectum) were computed within each corresponding organ-label map. The simulated dose map in GATE provides voxel-wise absorbed dose values in units of gray (Gy). In our case, since not all activities were simulated due to time limitations, the absorbed doses need to be normalized considering the simulated activity. Then the real average absorbed dose for a given organ $\bar{D_r}$ is calculated in the following way:

\begin{equation}
\bar{D_r} = \frac{1}{N_{\scriptscriptstyle V}} \sum_{v \in V} \left( \frac{D_{\scriptscriptstyle MC}}{A_{\scriptscriptstyle MC}} \cdot A_{ \scriptscriptstyle TIA} \right)
\end{equation}

\noindent
where $D_{\scriptscriptstyle MC}$ represents the Monte Carlo simulated absorbed dose (Gy) in each voxel $v$ within the organ volume $V$, as obtained from GATE. $A_{\scriptscriptstyle MC}$ denotes the activity (Bq) used in the MC simulation for each voxel $v$, and $A_{\scriptscriptstyle TIA}$ is the TIA (Bq) in the same voxel. $N_{V}$ indicates the total number of voxels in the organ. An example of MC  simulated dose image is shown in  Fig. \ref{fig:phantom}(d).

\subsection{Experiments and evaluation studies}

\subsubsection{Dataset preprocessing}

We conducted experiments on 1000 synthetic phantoms from the SynDoseTRT dataset, focusing on eight organs: prostate, salivary glands, liver, spleen, pancreas, kidneys, bladder, and rectum. The dataset was split into a training set (800 phantoms) and validation and test sets (100 phantoms each). The label consists of organ-wise dose values stored in comma-separated values (CSV) files.

To optimize memory usage, the network was designed for 2D image processing. 
Specifically, 3D PET images and the corresponding organ-label maps were projected along the coronal plane using voxel-wise intensity summation. Similar projection strategies have been applied in DL-based feature extraction from PET/CT images \cite{gil2023mip}. A third channel, representing the 2D mask of a specific organ index, was included to guide organ-level dose prediction. The resulting three-channel 2D input was used because the coronal projection preserves the global intensity distribution of PET, which may contain approximate mapping information related to organ-level dose values. 

\begin{figure}[t]
\centering
    \subfloat[Original PET]{\includegraphics[width=2.8cm]{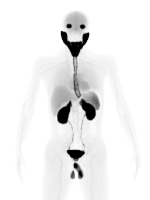}}~
    \subfloat[Weak augmentation]{\includegraphics[width=2.8cm]{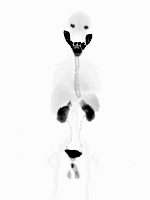}}~    
    \subfloat[Strong augmentation]{\includegraphics[width=2.8cm]{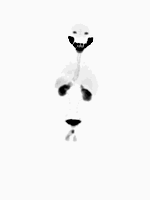}}~
    \caption{Example of data augmentation in SSL training: (a) original PET image; (b) weakly augmented image (rotation + Gaussian noise); (c) strongly augmented image (rotation + Gaussian noise + flipping + contrast etc.).}
    \label{fig:data_aug} 
\end{figure}

Data augmentation was applied on the training set. Following the common paradigm in SSL~\cite{sohn2020fixmatch-ssl}, we define two levels of augmentation: weak augmentation, which includes minimal spatial or noise perturbations, and strong augmentation, which introduces more diverse and intense transformations to challenge the model’s consistency. In our work, each image underwent weak augmentation (horizontal flipping, random rotation of ±5°, Gaussian noise) and strong augmentation (horizontal flipping, random rotation of ±5°, 50\% adjustment of brightness, contrast, and hue, Gaussian noise, and random perspective distortion with a scale of 0.5). Fig.~\ref{fig:data_aug} illustrated an example of weak and strong augmentation. In practice, we perform weak augmentation on labeled data, both weak and strong augmentation were performed on unlabeled data. In the proposed Pseudo-Label method, we generate ten versions of weakly augmented images per input. 

\subsubsection{Training protocol}

To assess the performance of the SSL approach, we trained various models using different proportions of labeled and unlabeled data. This was achieved by adjusting the ratio $\beta$ of labeled ($N_{\text{lbl}}$) to unlabeled ($N_{\text{ulb}}$) phantoms, as follows: 

\begin{equation}
N_{\text{ulb}} = (100 - \beta)\% \times N, \quad N_{\text{lbl}} = \beta\% \times N
\label{eq:nlabel}
\end{equation}

\noindent where $N$ is the total number of training images. The labeled ratio $\beta$ ranged from 5\% to 100\%. For each ratio $\beta$, we evaluated the performance on the test set (100 phantoms). The models are trained in two ways: (1) fully supervised learning (FSL), where $N_{\text{lbl}}$ were trained; (2) semi-supervised learning (SSL) using both $N_{\text{ulb}}$ and $N_{\text{lbl}}$.

Table \ref{tab:models} summarizes the SSL models, the parameter updating strategies and the hyperparameter settings during training for this study. The parameters ($\theta$) of the SSL models are updated through both supervised loss and unsupervised gradient updates. In order to validate the effectiveness of the proposed Pseudo-Label loss PSL method and the use of SSL in TRT, the SSL methods were evaluated and compared together. Note that in PSL method, pseudo-labels are generated without gradient updates to avoid reinforcing incorrect predictions. But the loss function computed between the pseudo-label and the model’s prediction is used for training and does contribute to gradient updates. We used two pre-trained models as backbone feature extractors in our SSL framework: ResNet50 \cite{he2016resnet} and Caformer\_s36 \cite{yu2024metaformer}, both obtained from the Timm library (v0.9.16) \cite{rw2019timm} and initially trained on ImageNet \cite{deng2009imagenet}.  ResNet50 uses skip (residual) connections that add a block’s input to its output, helping to prevent vanishing gradients and enabling deeper networks.  Caformer\_s36 integrates convolutional layers and Transformer-based attention \cite{vaswani2017attention}, enabling the model to effectively capture both local and global image features. The code was implemented with Python 3.10.14 and PyTorch 2.2.2 \cite{paszke2019pytorch}.

\begin{table}[t]
\centering
\caption{Parameter updating strategy of SSL models and hyperparameters configuration during training.}
\label{tab:models}
\resizebox{0.490\textwidth}{!}{%
\begin{threeparttable}
\begin{tabular}{llll}
\hline
Models &
  \begin{tabular}[c]{@{}l@{}}supervised part \end{tabular} &
  \begin{tabular}[c]{@{}l@{}}consistency reg.\end{tabular} &
  \begin{tabular}[c]{@{}l@{}}pseudo-label \end{tabular} \\ \hline
RegFixMatch (ours)  & with grad & with grad & without grad \\
MT \cite{tarvainen2017meanteacher-ssl} & with grad & EMA & without grad\\
ICT \cite{verma2022ict} & with grad & EMA &without grad\\
SGAN \cite{toutouh2023semigan} & with grad & with grad & without grad \\
SimRegMatch \cite{jo2024regssl} & with grad & N/A & without grad \\
\hline\hline
Hyperparamters\\\hline
Image shape & \multicolumn{3}{r} {256$\times$256$\times$3, 3 channels: attenuation map, PET, organ mask}\\
Backbone model & \multicolumn{3}{r} {ResNet50  \cite{he2016resnet} , Caformer\_s36 \cite{yu2024metaformer}}\\
Cross validation & \multicolumn{3}{r} {5-fold}\\
Epochs & \multicolumn{3}{r} {200}\\
Optimizer & \multicolumn{3}{r} {AdamW \cite{loshchilov2017adamw}, weight decay=1e-4}\\
Scheduler &\multicolumn{3}{r} {ExponentialLR}\\
Learning rate & \multicolumn{3}{r} {1e-4}\\
Batch size & \multicolumn{3}{r} {10}\\
Unsup. \& sup. loss  & \multicolumn{3}{r}  {MAE}\\
Unsup. loss weight $\alpha$ & \multicolumn{3}{r}  {0.1}\\
Metrics & \multicolumn{3}{r}  {$R^2$ score, MAPE, PCC}\\

\hline
\end{tabular}
\end{threeparttable}
}
\end{table}

\subsubsection{Evaluation Metrics}

The model performance was assessed using three evaluation metrics, which are calculated based on organ-wise dose. Among these, the Mean Absolute Percentage Error (MAPE) quantifies the relative error between the predicted dose ($D_{\text{pred}}$) and the MC dose ($D_{\text{MC}}$) as reference or label, according to the following formula: 

\begin{equation}
    \mathrm{MAPE}(D_{\text{pred}}, D_{\text{MC}}) = \frac{|D_{\text{pred}} - D_{\text{MC}}|}{D_{\text{MC}}} \times 100\%
    \label{eq:mape}
\end{equation}

Coefficient of Determination ($R^2$ Score) evaluates how well the model explains variance in the dose values, which is defined as below:

\begin{equation}
R^2 = 1 - \frac{\sum (D_{\text{MC}} - D_{\text{pred}})^2}{\sum (D_{\text{MC}} - \bar{D}_{\text{MC}})^2}
\label{eq:r2}
\end{equation}

\noindent where the numerator represents the sum of squared residuals, which quantifies the discrepancy between the predicted and reference doses. The denominator represents the total sum of squares, measuring deviations from the mean MC dose $\bar{D}_{\text{MC}}$. The $R^2$ score ranges from $-\infty$ to 1, with values closer to 1 indicating better model fit and negative values suggesting performance worse than using the mean dose as a predictor.

As well as Pearson Correlation Coefficient (PCC), the PCC measures the linear correlation between predicted and MC dose values:

\begin{equation}
    \rho(D_{\text{pred}}, D_{\text{MC}}) = \frac{\mathrm{cov}(D_{\text{pred}}, D_{\text{MC}})}{\sigma_{D_{\text{pred}}} \sigma_{D_{\text{MC}}}}
    \label{eq:rho}
\end{equation}

\noindent where $\mathrm{cov}(D_{\text{pred}}, D_{\text{MC}})$ is the covariance between predicted and reference doses, $\sigma_{D_{\text{pred}}}$ and $\sigma_{D_{\text{MC}}}$ are their standard deviations. The PCC ranges from $-1$ to $1$, where values close to 1 indicate a strong positive correlation, values near 0 suggest no correlation, and negative values imply an inverse relationship.

\subsection{Sensitivity of the personalized dose prediction}

Lastly, we investigated the impact of individual variations in organ-specific radioactive activity. Since a considerable portion of the parameters in PBPK model (see Eq. \ref{eq:pbpk}) rely on population-averaged physiological and biochemical parameters derived from literature or clinical statistics. This may limit its accuracy in personalized dose predictions for individual patients. On the other hand, the proposed SSL model is a data-driven learning method designed to learn complex patterns and relationships directly from training data. This enables it to capture individual-specific variations in organ dose distributions. By leveraging both labeled and unlabeled data, the SSL model can adapt to the unique characteristics of each individual, making it suitable for personalized dose prediction. 

To evaluate the performance of the SSL model in this context, we introduced subtle and random variations in organ-specific activity ranging from -2\% to 2\% of the original values generated by the PBPK model in each phantom of the data set. This mimic the variability in radioactive activity that may occur across individuals due to differences in physiology, metabolism, or other factors. Correspondingly, the PET images and dose maps were also updated.

\section{Results and discussions}




\subsection{Performance of SSL models}

\begin{figure}[t]
    \centering
    \includegraphics[width=8.0cm]{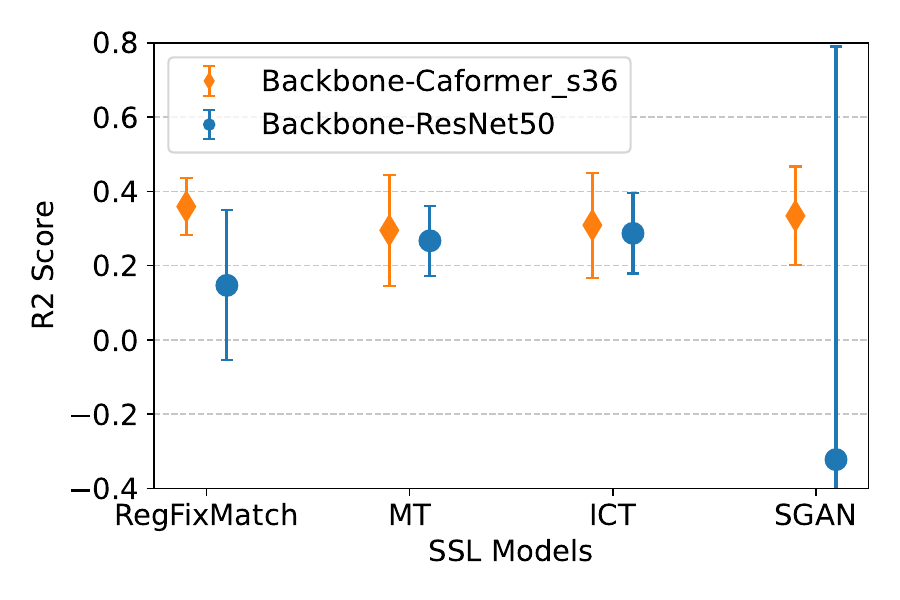}
    \caption{$R^2$ score ($\uparrow$) of SSL models with backbones of ResNet50 and Caformer\_s36 respectively.}
    \label{fig:plot-bar-backbone}
\end{figure}

The choice of backbone model can impact SSL performance. From Fig. \ref{fig:plot-bar-backbone} we observed that Caformer\_s36 consistently yielded higher $R^2$ scores across RegFixMatch, MT, and ICT methods than ResNet50, with particularly strong gains for SGAN. Besides, Caformer\_s36 exhibited lower performance variability (smaller error bars) in RegFixMatch and SGAN models, indicating greater stability. This enhanced performance and stability likely stem from attention mechanism of Caformer\_s36. This mechanism allows the model to better capture global context and long-range dependencies within the data, features potentially crucial for regression tasks, compared to the more locally focused convolutional operations inherent in ResNet50. Based on the above study results, in the subsequent experimental comparisons, unless otherwise specified, the SSL models adopt Caformer\_s36 as the backbone and incorporate PSL to boost the performance.

\begin{figure}[t]
\centering
    \subfloat[RegFixMatch (Proposed)]{\includegraphics[width=4.40cm]{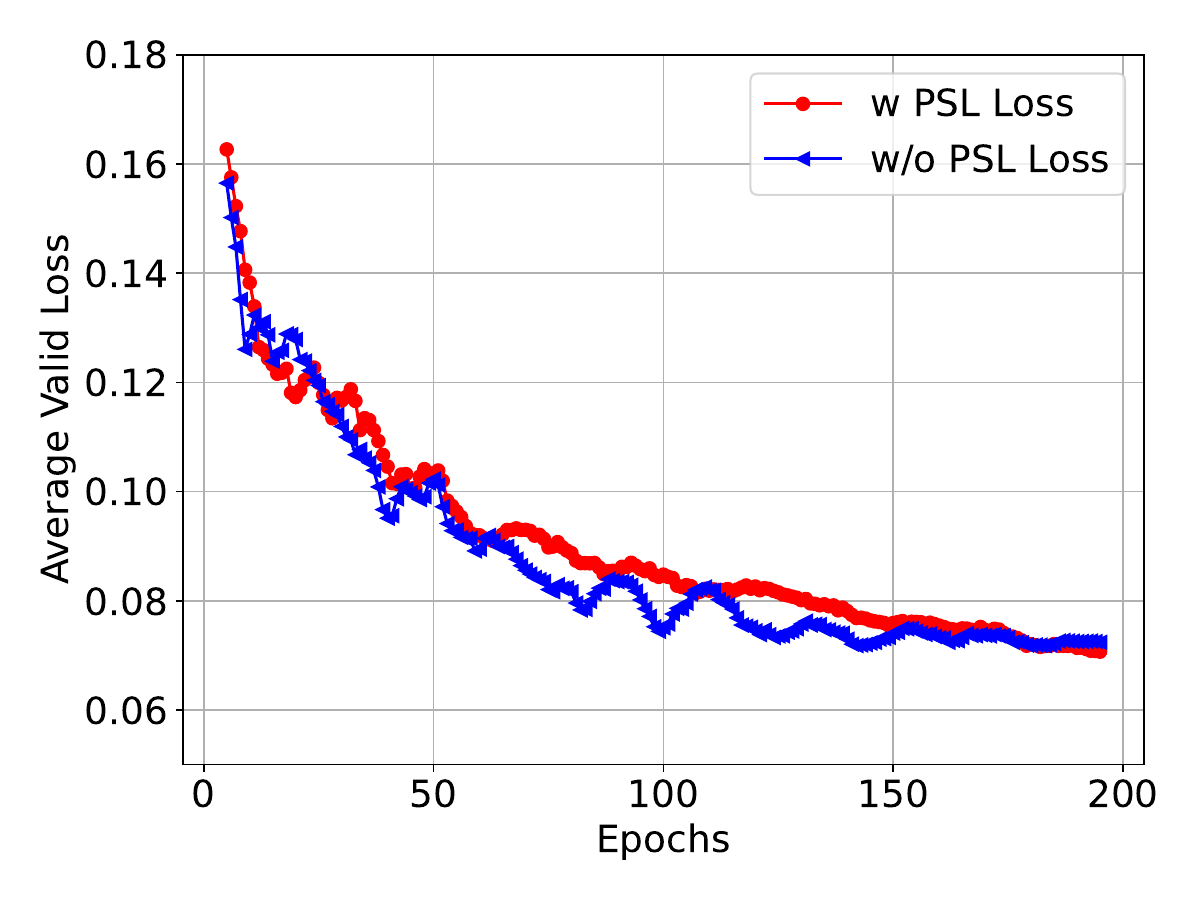}}
    \subfloat[MT]{\includegraphics[width=4.40cm]{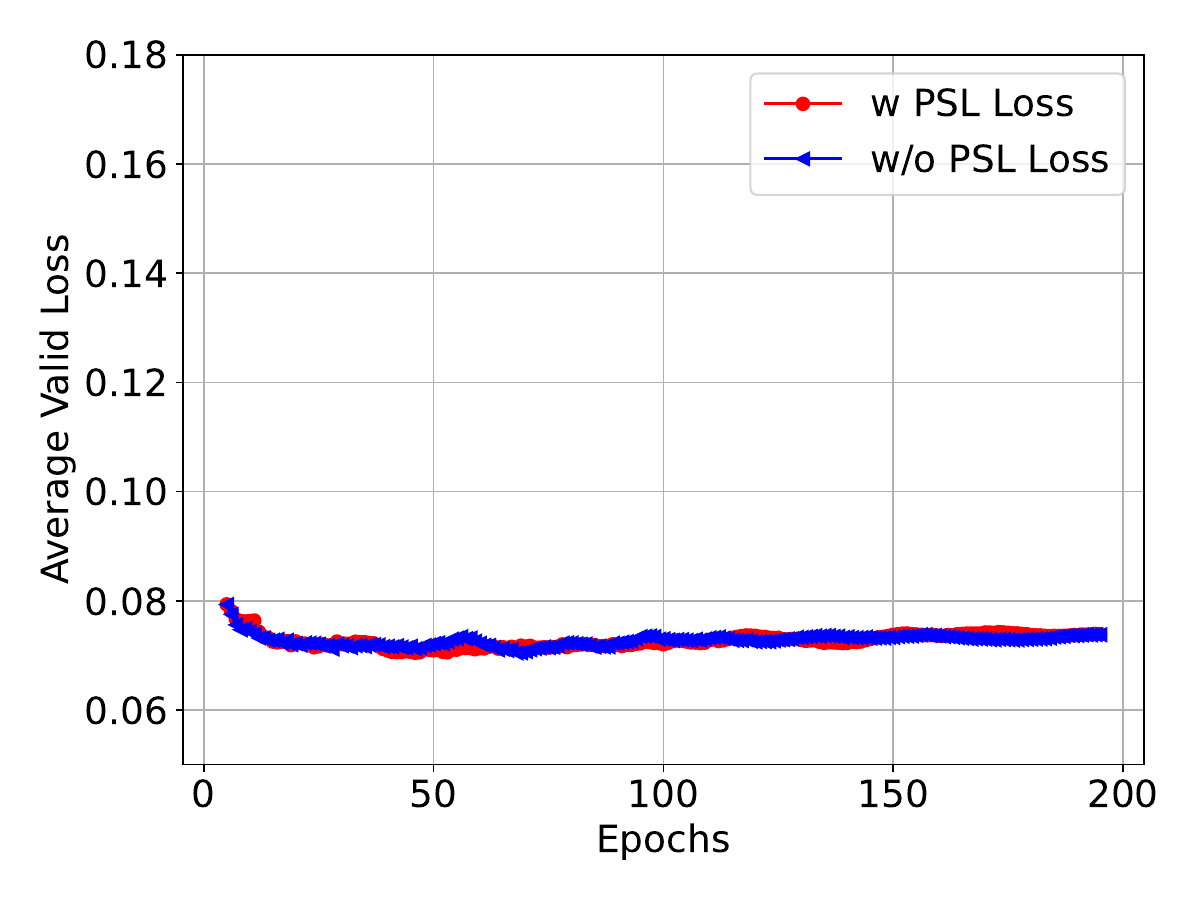}}
    
    \subfloat[ICT]{\includegraphics[width=4.40cm]{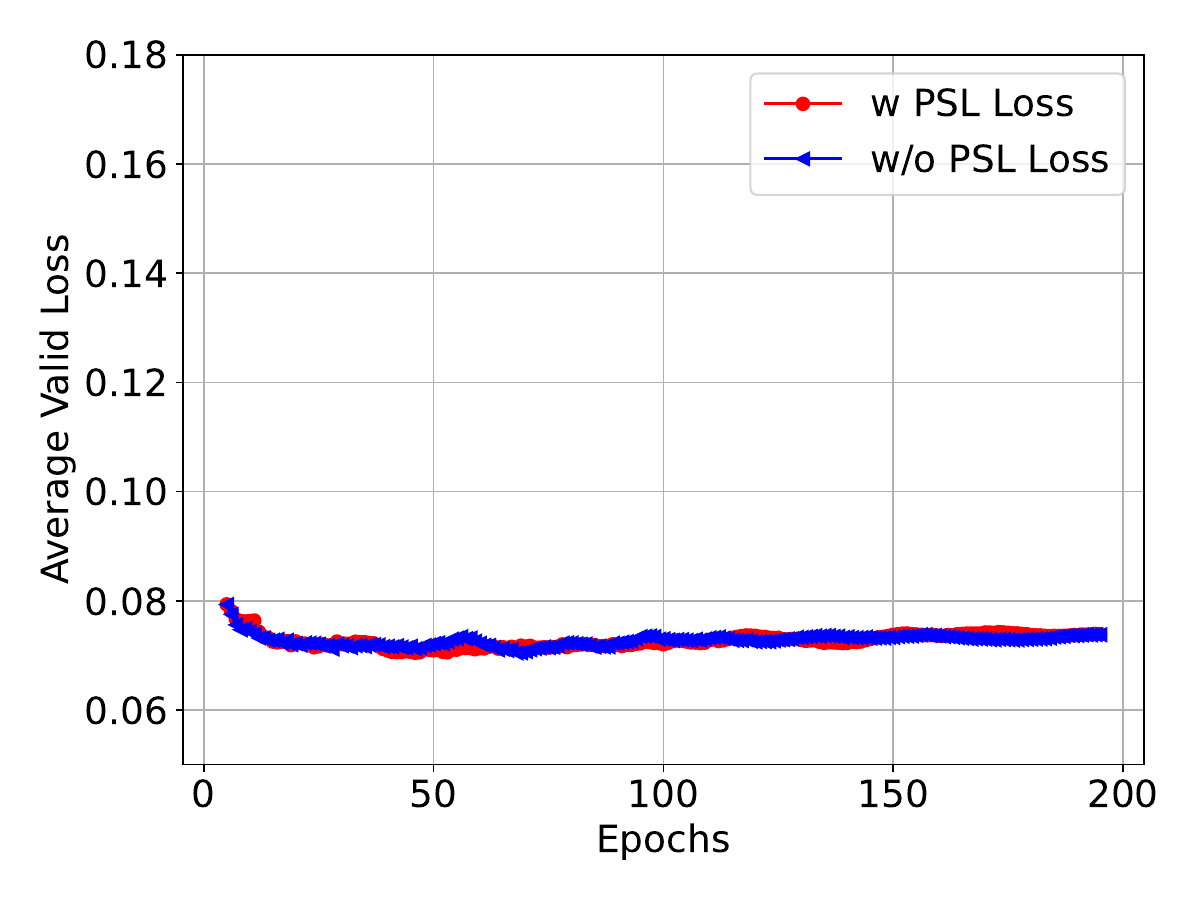}}
    \subfloat[SGAN]{\includegraphics[width=4.40cm]{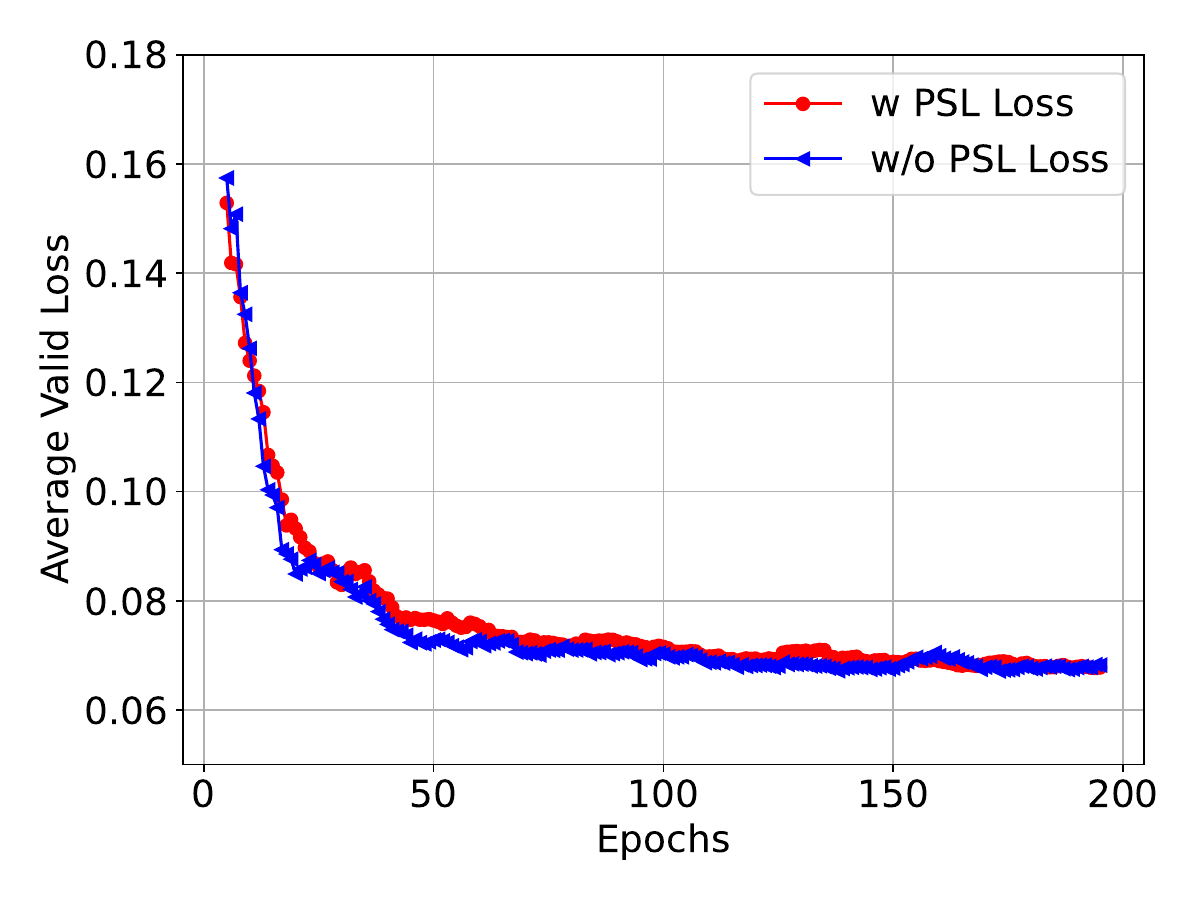}}
    \caption{Average valid loss curves over 5-fold cross-validation training of different SSL models for $\beta=25\%$ of labeled data with and without Pseudo-Label loss.}
    \label{fig:loss-curves} 
\end{figure}

We analyzed the validation-stage learning curves of the SSL models for the labeled rate $\beta=25\%$. Results are shown in Fig. \ref{fig:loss-curves}. Overall, none of these models exhibited overfitting or underfitting, suggesting that they effectively learned meaningful representations from the augmented dataset. Specifically, one can observe that the proposed RegFixMatch model with PSL achieved lower validation loss (and potentially continue to descend with more training epochs) and smoother convergence compared to the model without PSL and other SSL models. The possible reason is that the PSL provides additional supervision, reducing uncertainty during the learning process and guiding the model toward more stable optimization. In contrast, the learning curves of the SGAN model decreased quickly, particularly in the early stages. This rapid decline may be attributed to its architecture, where the interplay between the generator and discriminator enhances feature learning and accelerates convergence during training. Meanwhile, the learning curves of MT and ICT exhibited a flatter trend with no further loss reduction. This behavior may be linked to their parameter update mechanisms (see Table \ref{tab:models}), the supervised components are updated via the gradient based i.e. AdamW optimizer, while EMA in the unsupervised parts slows updates, stabilizing rather than optimizing the loss further.

\begin{figure}[t]
     \centering
     \includegraphics[width=7.00cm]{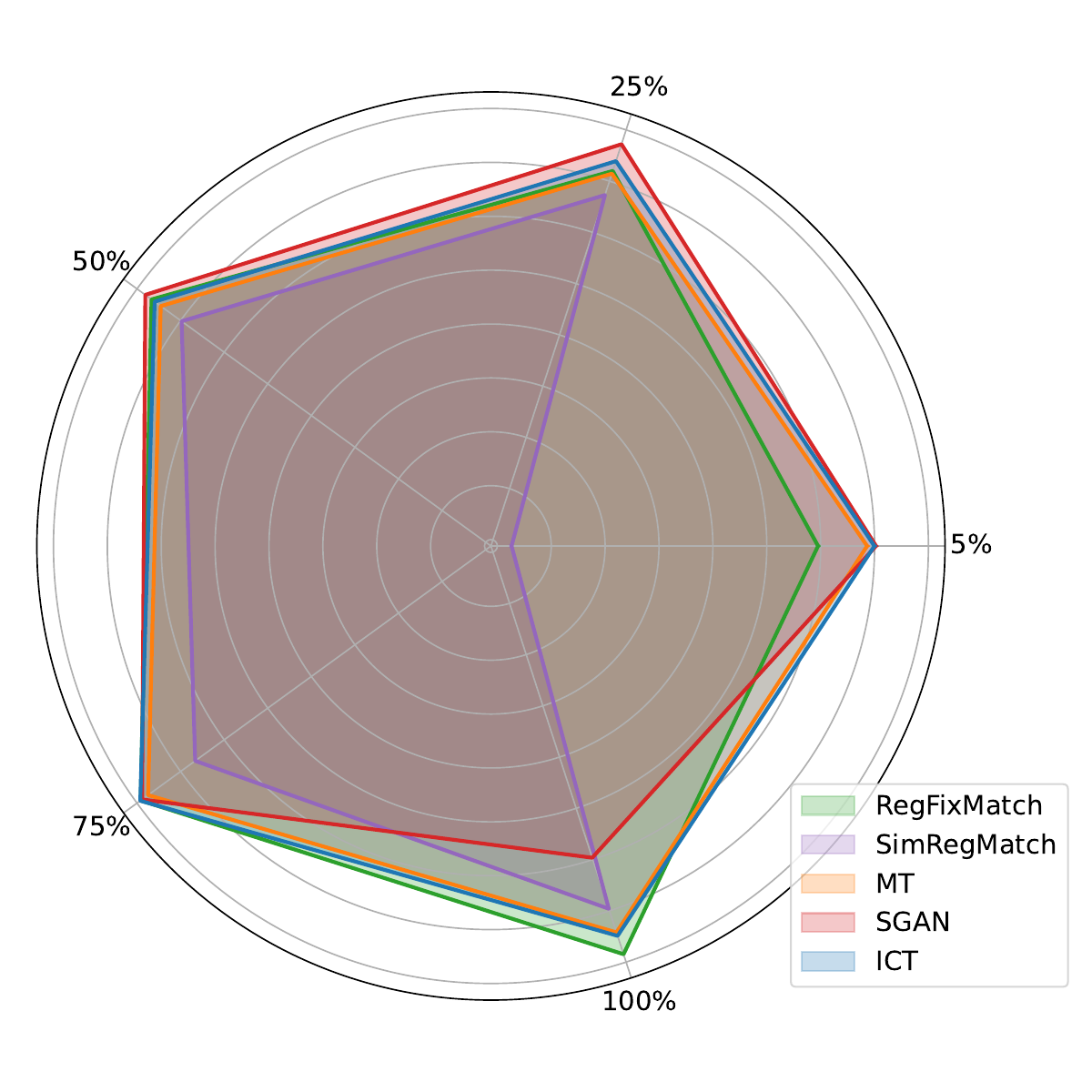}
     \caption{Radar chart of the average performance over all organs of SSL models with different labeled rates ($\beta \in \{5\%, 25\%, 50\%, 75\%, 100\%\}$) of training set.}
     \label{fig:radar-models}
 \end{figure}
 
A radar chart (Fig. \ref{fig:radar-models}) was used to visualize the overall performance of the SSL models with different proportions of labeled data. The proposed RegFixMatch demonstrated stable improvements as the amount of labeled data increased. SGAN with the proposed PSL outperformed other SSL models, which may due to its integration of generative adversarial networks with a semi-supervised learning strategy. The inclusion of a generator allowed the model to synthesize additional training samples, enhancing data diversity and improving generalization, particularly when labeled data is limited. This generative component provided an auxiliary supervisory signal, helping the discriminator learn more robust feature representations. The loss function designed in SGAN further contributed to its effectiveness, as it combined supervised loss, consistency loss, adversarial loss, and Pseudo-Label loss, ensuring stable training and improved performance across different data proportions. ICT and MT showed moderate performance across all settings. SimRegMatch, on the other hand, was less competitive across different labeled data proportions. 

We further studied SimRegMatch\cite{jo2024regssl}, which is the closest method to the proposed RegFixMatch (FixMatch with the proposed PSL). SimRegMatch uses a pseudo-label calibration strategy based on uncertainty limitation and incorporates feature similarity through random weight dropout\cite{srivastava2014dropout}. This operation could cause significant memory overhead during training.  Table \ref{tab:memory} presents this comparison. SimRegMatch allocated 25,312 Megabytes (MB) of memory, whereas the proposed method required only 563 MB. The results indicate that the proposed method not only reduced memory usage during training but also achieved superior performance. Specifically, the average $R^2$ score across four labeled data ratios ($\beta \in \{5\%,25\%,50\%,75\%\}$) for RegFixMatch was higher (58.2\%), with a value of $0.174\pm0.072$, compared to SimRegMatch, which had a value of $0.110\pm0.312$.

\begin{table}[t]
\centering
\caption{Memory usage (MB) of SimRegMatch and our proposed method RegFixMatch. Backbone is ResNet50.}
\label{tab:memory}
\resizebox{0.49\textwidth}{!}{%
\begin{tabular}{lrr}
\hline
  & SimRegMatch  &  RegFixMatch \\ \hline
Nb of parameters in model & 23.5  & 23.5 \\
Model memory &  89.9  & 89.9 \\
Allocated memory & 12 169  & 563 \\
Reserved memory & 25 312  &  5 550\\
\hline
\end{tabular}
}
\end{table}

\begin{figure}[t]
    \centering
    \includegraphics[width=8.8cm]{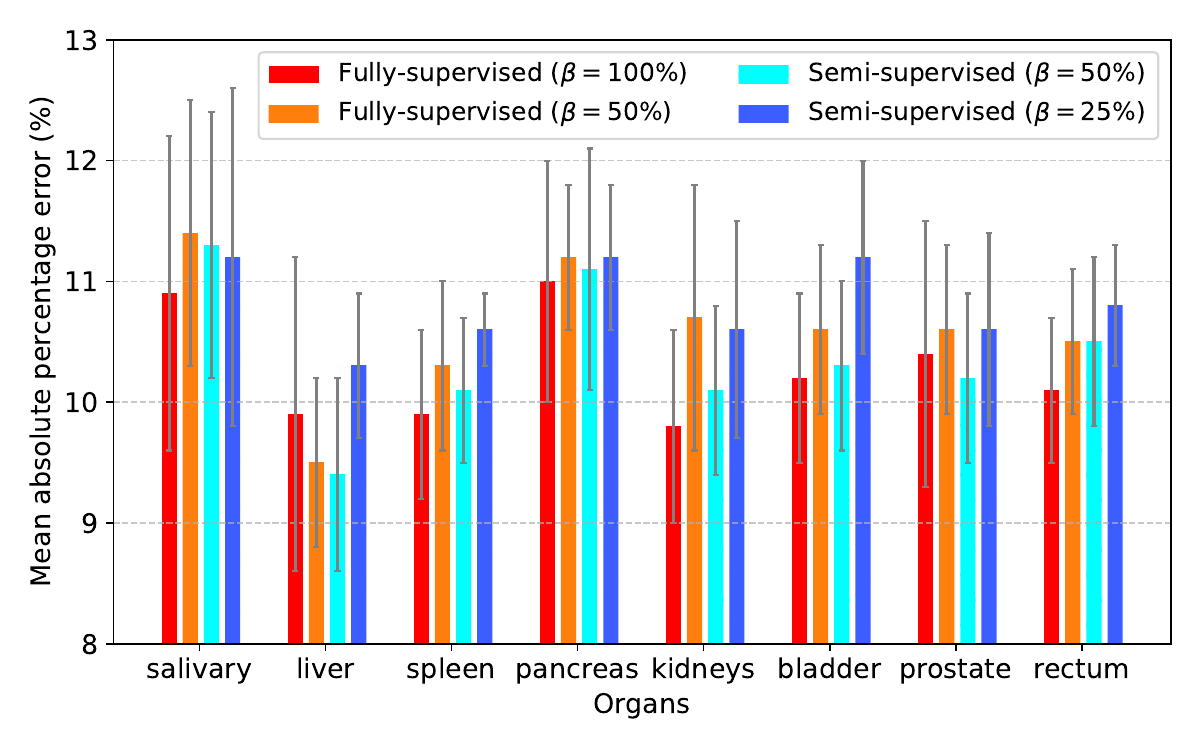}
    \caption{Dose prediction error (MAPE) respectively on FSL models ($\beta=100\%$ and $\beta=50\%$) and on SSL models (RegFixMatch) with $\beta=50\%$ and $\beta=25\%$ labeled data of different organs.}
    \label{fig:fsl-ssl}
\end{figure}

\subsection{Dose prediction with FSL vs. SSL}

\begin{table*}
\centering
\caption{Comparative analysis of organ dose prediction (taking kidneys as an example) with state-of-the-art semi-supervised learning (SSL) methods using the SynDoseTRT Phantom dataset with different ratios of labeled data. The gray colored row is our proposed Pseudo-Label loss (PSL). The evaluation metrics specifically $R^2$ score, MAPE and PCC assesses average performance and standard deviation ($\pm$), over 5-fold cross validation. The highest-performing method is highlighted in bold.}
\label{tab:results}
\tiny
\renewcommand{\arraystretch}{0.5} 
\resizebox{\textwidth}{!}{
\begin{tabular}{lccccc}

\toprule
\textbf{SSL methods} & \textbf{Labeled rate} & $\beta=5\%$ & $\beta=25\%$ & $\beta=50\%$ & $\beta=75\%$  \\
\midrule
\multicolumn{6}{c}{MAPE $\downarrow$} \\
\midrule
\multirow{3}{*}{MT}  & Supervised & $10.6\%\pm0.7$ & $10.4\%\pm0.8$ & $10.3\%\pm0.7$ & $10.2\%\pm0.5$ \\
& w/o PSL  & $10.6\%\pm0.3$ & $10.6\%\pm0.8$ & $10.2\%\pm0.6$ & $10.1\%\pm0.4$ \\
\rowcolor{gray!20}
& w/ PSL & $\textbf{10.4}\%\pm\textbf{0.4}$ & $10.7\%\pm0.9$ & $10.6\%\pm0.7$ & $10.3\%\pm0.4$ \\
\midrule
\multirow{3}{*}{ICT} & Supervised &  $10.8\%\pm0.6$ & $10.6\pm0.5$ & $10.3\%\pm0.8$ & $10.4\%\pm0.4$ \\
& w/o PSL & $10.7\%\pm0.9$ & $10.3\%\pm0.8$ & $10.4\%\pm0.4$ & $10.1\%\pm0.4$ \\
\rowcolor{gray!20}
& w/ PSL & $10.8\%\pm0.7$ & $\textbf{10.1}\%\pm\textbf{0.7}$ & $10.3\%\pm0.6$ & $10.5\%\pm0.7$ \\
\midrule
\multirow{3}{*}{SGAN} & Supervised & $10.8\%\pm1.0$ & $10.3\%\pm0.6$ & $10.2\%\pm0.6$ & $10.3\%\pm0.8$  \\
& w/o PSL & $10.9\%\pm0.7$ & $10.3\%\pm0.9$& $10.3\%\pm0.7$ & $10.4\%\pm0.4$ \\
\rowcolor{gray!20}
& w/ PSL & $10.5\%\pm0.6$ & $10.3\%\pm0.8$ & $10.3\%\pm0.2$ & $10.1\pm0.7$ \\
\midrule
\multirow{3}{*}{RegFixMatch} & Supervised & $11.5\%\pm0.7$ & $10.8\%\pm0.6$ & $10.7\%\pm1.1$ & $10.2\%\pm0.6$ \\
& w/o PSL & $11.6\%\pm0.5$ & $10.5\%\pm0.5$& $10.3\%\pm0.6$ & $\textbf{10.0}\%\pm\textbf{0.8}$ \\
\rowcolor{gray!20}
& w/ PSL & $11.6\%\pm0.6$ & $10.6\%\pm0.9$ & $\textbf{10.1}\%\pm\textbf{0.7}$ & $10.2\%\pm0.7$ \\
\midrule
\multicolumn{6}{c}{$R^2$ score $\uparrow$} \\
\midrule
\multirow{3}{*}{MT}  & Supervised & $0.148\pm0.134$ & $0.149\pm0.192$ & $0.193\pm0.122$ & $0.225\pm0.099$ \\
& w/o PSL & $0.142\pm0.132$ & $0.119\pm0.189$ & $0.196\pm0.072 $ & $0.215\pm0.068 $  \\
\rowcolor{gray!20}
& w/ PSL & $\textbf{0.172}\pm\textbf{0.148}$ & $0.123\pm0.198$ & $0.151\pm0.153$ & $0.211\pm0.104$ \\
\midrule
\multirow{3}{*}{ICT}  & Supervised & $0.127\pm0.109$& $0.132\pm0.145$ & $0.194\pm0.091$ & $0.194\pm0.103$ \\
& w/o PSL & $0.130\pm0.138$ & $0.169\pm0.163$ & $0.176\pm0.092$ & $0.237\pm0.095$  \\
\rowcolor{gray!20}
& w/ PSL & $0.116\pm0.131$ & $\textbf{0.218}\pm\textbf{0.111}$ & $0.201\pm0.086$ & $0.184\pm0.111$ \\
\midrule
\multirow{3}{*}{SGAN} & Supervised & $0.072\pm0.270$& $0.186\pm0.114$ & $\textbf{0.206}\pm\textbf{0.082}$ & $0.206\pm0.045$ \\
& w/o PSL & $0.090\pm0.110$ & $0.197\pm0.129$ & $0.178\pm0.141$ & $0.187\pm0.065$ \\
\rowcolor{gray!20}
& w/ PSL & $0.167\pm0.093$ & $0.187\pm0.098$ & $0.187\pm0.117$ & $0.218\pm0.068$  \\
\midrule
\multirow{3}{*}{RegFixMatch} & Supervised & $0.017\pm0.099$ & $0.124\pm0.128$ & $0.138\pm0.101$ & $0.217\pm0.044$  \\
& w/o PSL & $-0.012\pm0.016$ & $0.132\pm0.123$ & $0.182\pm0.097$ & $\textbf{0.242}\pm\textbf{0.089}$  \\
\rowcolor{gray!20}
& w/ PSL & $-0.011\pm0.014$ & $0.131\pm0.171$ & $0.191\pm0.093$ & $0.216\pm0.075$ \\
\midrule
\multicolumn{6}{c}{PCC $\uparrow$} \\
\midrule
\multirow{3}{*}{MT} & Supervised & $0.403\pm0.139$ & $0.434\pm0.160$ & $0.470\pm0.098$ & $0.496\pm0.085$ \\
& w/o PSL & $0.401\pm0.147$ & $0.426\pm0.140  $ & $\textbf{0.481}\pm\textbf{0.076}$ & $0.481\pm0.064$ \\
\rowcolor{gray!20}
& w/ PSL & $\textbf{0.430}\pm\textbf{0.145}$ & $0.424\pm0.147$ & $0.466\pm0.077$ & $0.489\pm0.091$  \\
\midrule
\multirow{3}{*}{ICT} & Supervised & $0.392\pm0.105$ & $0.432\pm0.099$ & $0.462\pm0.081$ & $0.468\pm0.106$ \\
& w/o PSL & $0.401\pm0.126$ & $0.436\pm0.130$ & $0.447\pm0.080$ & $\textbf{0.506}\pm\textbf{0.073}$  \\
\rowcolor{gray!20}
& w/ PSL & $0.395\pm0.118$ & $\textbf{0.466}\pm\textbf{0.118}$ & $0.463\pm0.079$ & $0.465\pm0.091$  \\
\midrule
\multirow{3}{*}{SGAN} & Supervised & $0.368\pm0.179$ & $0.438\pm0.125$ & $0.472\pm0.066$ & $0.474\pm0.071$ \\
& w/o PSL & $0.313\pm0.195$ & $0.454\pm0.125$ & $0.468\pm0.103$ & $0.457\pm0.077$ \\
\rowcolor{gray!20}
& w/ PSL & $0.424\pm0.103$ & $0.446\pm0.107$ & $0.464\pm0.089$ & $0.491\pm0.085$  \\
\midrule
\multirow{3}{*}{RegFixMatch} & Supervised & $0.248\pm0.144$ & $0.378\pm0.133$ & $0.437\pm0.062$ & $0.487\pm0.050$ \\
& w/o PSL & $-0.034\pm0.317$ & $0.409\pm0.134$ & $0.455\pm0.095$ & $0.503\pm0.084$ \\
\rowcolor{gray!20}
& w/ PSL & $0.030\pm0.588$ & $0.421\pm0.136$ & $0.457\pm0.082$ & $0.489\pm0.068$  \\
\bottomrule
\end{tabular}
}
\end{table*}

A comparative analysis of organ dose prediction across the SSL models and varying labeled data ratios ($\beta$), evaluated with and without the proposed PSL was summarized in Table \ref{tab:results}. Overall, the results suggested that incorporating PSL consistently boosted or at least maintained performance across most settings. For example, when only 5\% of the data was labeled, PSL-embeded MT approach achieve an average MAPE of 10.4\% (10.6\% in fully supervised way). When 25\% of the data was labeled, the model ICT with proposed PSL had lowest MAPE score (10.1\%) among other SSL methods, which is also lower than fully supervised approach (10.6\%). When 50\% and 75\% of the data was labeled, the proposed RegFixMatch method performed better (MAPE is 10.1\% and 10.0\%) than other methods including fully supervised methods. Similar results with the PCC score. This suggested that different SSL methods tend to excel under different labeling scenarios, highlighting the importance of selecting the appropriate method based on the amount of labeled data available.  Of particular note is that PSL’s benefits diminished slightly in high labeled-data regimes (e.g., 75\% labeled ratio), where RegFixMatch with PSL exhibited a marginal $R^2$ decline (0.216 vs. 0.242 without PSL). This suggested potential noise amplification from pseudo labels in labeled-data-rich scenarios. Although certain subsets of results particularly at higher labeling ratios showed less pronounced gains, the majority of the comparisons indicated that proposed PSL provided robust advantages in SSL models. Beyond that, this Pseudo-Label loss can be readily integrated into diverse semi-supervised learning frameworks.

The peformance of the proposed RegFixMatch was compared against fully supervised learning (FSL). The dose prediction error rate (MAPE) was around 9\%-11\%, as illustrated in Fig. \ref{fig:fsl-ssl}, showing the capability of DL methods to predict dose with acceptable accuracy. The proportion of labeled data significantly impacted prediction accuracy: as the fraction of labeled data decreased, MAPE increased accordingly, which aligns with expectations. Nevertheless, the SSL approach demonstrated clear advantages. Specifically, when 50\% of the data was labeled, the SSL model achieved for some organs a prediction error comparable to the FSL model trained on 100\%  labeled data. For example the MAPE value for the spleen was 9.8\% and 10.1\% for FSL and SSL respectively. Notably, a comparison between SSL and FSL under the same proportion of labeled data ($\beta=50\%$) revealed that the dose prediction error rates of SSL were lower than FSL in most organs, see Fig. \ref{fig:fsl-ssl}. For instance, the kidney dose prediction error rate of FSL was 10.7\%, whereas it was 10.1\% in SSL. The only exception was the rectum, where both methods yielded same error rates of 10.5\%. This advantage of SSL can be attributed to its ability to leverage additional unlabeled data to learn the pattern from them, thereby enhancing generalization performance even with limited labeled samples. Therefore, in the case of insufficient labeled data, the strategy of semi-supervised learning to make full use of unlabeled data learning can be a priority consideration.

\subsection{Sensitivity of the personalized dose prediction}

\begin{figure}[t]
    \centering
    \includegraphics[width=9cm]{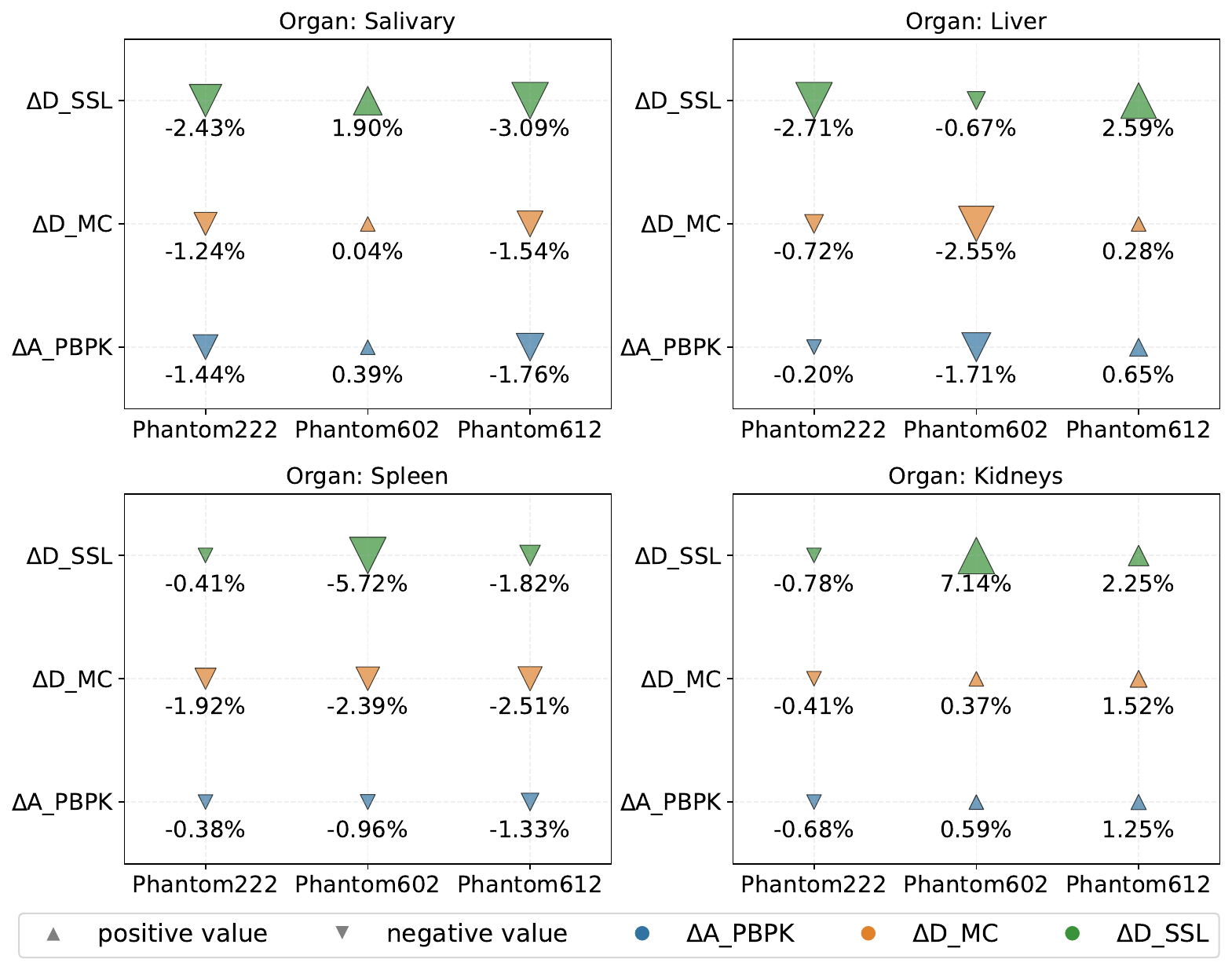}
    \caption{Comparison of changes in organ dose (D) predictions based on Monte Carlo (MC) and SSL methods, respectively, brought by individual variation in radioactive activity (A) generated by the PBPK model.}
    \label{fig:vardose}
\end{figure}

We selected four organs from three random phantoms in the test set as examples and observed the changes in activity, MC dose, and SSL predicted dose using the RegFixMatch. Results are presented in Fig. \ref{fig:vardose} and show that the most organ predicted dose from SSL follows the variation of the activity within the different organs. For instance, the salivary glands experienced a dose variation of -1.24\% according to MC simulation and -2.43\% according to SSL prediction, indicating that the SSL model accurately captured the dose variation patterns within the pretherapeutic PET image. Yet exceptions happened to a small part of organs that the variations of MC dose and SSL dose were not consistent, for example, in the kidneys of phantom 602, the MC dose value increased by 0.37\%, while the SSL dose decreased by 7.14\%. These uncertainties may be attributed to limitations in the data size and diversity or model architectures, and this could be controlled by expanding training data (current dataset has 1000 phantoms) or incorporating physical constraints into the modeling process.

In addition to the consistency of variations, we also evaluated the dose prediction error on the original and varied data set respectively, see Table \ref{tab:vardose}. The MAPE results for the original and varied dose predictions across different organs revealed that the SSL model maintained stable predictive accuracy despite variations in organ doses, which indicated the learning capacity of SSL models. For instance, the original prediction error of the liver was 8.5\%, and after dose variation, it was 8.8\%. This minimal change in error rates underscored the robustness of the SSL model. 

\begin{table}[ht]
\centering
\caption{MAPE on the test set and varied test set.}
\label{tab:vardose}
\renewcommand{\arraystretch}{1.5} 

\resizebox{0.5\textwidth}{!}{%
\begin{tabular}{lcccccccc}
\hline
 Organs &  Salivary & Liver & Spleen  & Pancreas & Kidneys& Bladder& Prostate& Rectum \\ \hline
MAPE\_ori & 10.7\% & 8.5\% & 10.1\% & 9.8\% & 10.5\% & 11.1\% & 9.3\% & 11.3\%    \\
MAPE\_var & 10.1\% & 8.8\% & 10.5\% & 9.5\% & 10.9\% & 11.1\% & 9.4\% & 11.5\%   \\
\hline
\end{tabular}
}
\end{table} 

\section{Conclusions}

In this work, we presented semi-supervised learning for dose prediction in targeted radionuclide therapy, the first study that predicting personalized doses from pre-therapy PET images with limited labeled dose data. The study also introduced SynDoseTRT, a synthetic phantom dataset based on PBPK models and Monte Carlo simulations, which can serve as a resource for AI-based dosimetry studies. While the proposed SSL framework shows potential, some limitations were observed: the addition of Pseudo-Label loss did not consistently improve dose prediction across different models and labeling rates on all the organs listed in this experiment. Moreover, even under promising labeling scenarios, the final prediction error remains limited (approximately 10\% error), this could due to 2D projection images and insufficient data scale. Future work will focus on exploring full 3D representations to better capture organ-level features, and further enriching the SynDoseTRT dataset with more realistic characteristics such as tumors, as well as reducing dependency on labeled data of the SSL models through network capacity optimization or self-supervised learning strategies. Ultimately, our goal is to establish a flexible, standardized platform capable of simulating patient-specific dose distributions, which can be applied to real clinical data to support diagnosis and treatment planning.

\section*{CONFLICT OF INTEREST}
All authors declare that they have no known conflicts of interest in terms of competing financial interests or personal relationships that could have an influence or are relevant to the work reported in this paper. 

\bibliographystyle{unsrt}
\bibliography{reference.bib} 
\end{document}